\documentstyle[12pt]{article}
\newcommand{\doublespace}{\renewcommand{\baselinestretch}{1.75}
\Large\normalsize}

\begin{document}
\doublespace
\begin{titlepage}

\centerline{\bf SPACE-TIME DEFECTS AND TELEPARALLELISM} 
\vskip 1.0cm   
\bigskip
\centerline{\it J. W. Maluf$\,^{*}$ and A. Goya}
\centerline{\it Instituto de F\'isica}
\centerline{\it Universidade de Bras\'ilia}
\centerline{\it C.P. 04385}
\centerline{\it 70.919-970  Bras\'ilia, DF}  
\centerline{\it Brazil}
\date{}
\begin{abstract}
We consider the class of space-time defects  investigated
by Puntigam and Soleng. These defects describe space-time
dislocations and disclinations (cosmic strings), and are in
close correspondence to the actual defects that arise in crystals
and metals. It is known that in such materials dislocations
and disclinations require a small and large amount of energy,
respectively, to be created. The present analysis is carried out
in the context of the teleparallel equivalent of general
relativity (TEGR). We evaluate the gravitational energy of these
space-time defects in the framework of the TEGR and find that
there is an analogy between defects in space-time and in continuum
material systems: the total gravitational energy of space-time
dislocations and disclinations (considered as idealized defects)
is zero and infinit, respectively.

\end{abstract}
\thispagestyle{empty}
\vfill
\par
\noindent (*) e-mail: wadih@fis.unb.br
\end{titlepage}
\newpage

\noindent {\bf I. Introduction}\par
\bigskip
\noindent Space-time defects are assumed to play an important 
role in the large-scale structure of the universe. The cosmic 
strings are prominent examples. The belief is that 
they may  have acted as seeds in the process of formation
of galaxies and cluster of galaxies, hence contributing
to the problem of the origin of the initial density
fluctuations\cite{Vilenkin}. The two kinds of space-time
defects, dislocations and disclinations (cosmic strings), 
are geometrical constructions that stem from actual 
defects that occur in crystals and metals\cite{Chaikin}.
A discussion of the differential geometry and
topology of these defects, in the context of condensed 
matter physics, is given in Ref. \cite{Rivier}, and 
a pictorial exposition in Ref. \cite{Harris}. The similarity
between some gravitational field configurations and physical 
realizations of material systems like crystals and metals is
by itself a very interesting feature.

Topological defects like dislocations and
disclinations, which are also called {\it Volterra
distortions}, have recently been considered in the literature in
the context of the Einstein-Cartan theory\cite{Puntigam}. A
systematic investigation  of the analogy between distortions of
solids and defect structures in Riemann-Cartan manifolds has been
carried out in the latter reference. 
One outcome of this investigation is the classification
of dislocations and disclinations in Riemann-Cartan space-times.
Since the Einstein-Cartan field equations considered in
Ref. \cite{Puntigam} have an effective Einsteinian
form\cite{Puntigam,Hehl1}, these Volterra distortions are also
defect structures in Einstein's general relativity.

In the framework of general relativity
space-time defects can be characterized by metric functions that,
at least in the examples considered here, can be 
transformed into a flat space-time metric  away from the
defect axis. It is then concluded that the defect is concentrated
in the $z$ axis and that it is mathematically realized as
delta functions with support in the axis. Such 
interpretation is given, for instance, in 
Refs. \cite{Puntigam,Galtsov,Tod,Letelier}. The energy-momentum
tensors that generate  simple space-time dislocations and 
disclinations have been
computed in Refs. \cite{Galtsov,Letelier},
in the case where the defect parameters are constants. In the
context of Einstein's equations it is concluded that the
corresponding  energy-momentum tensors contain linear and
quadratic terms in the two-dimensional delta function. The
quadratic terms are simply discarded, and eventually not
considered as anomalies of the theory. The final structure of
the energy-momentum tensors is then taken to support the above
interpretation regarding the localizability of the defect.

In this paper we will address space-time defects in the
context of the teleparallel equivalent of general relativity
(TEGR)\cite{Moller,Hehl2,Maluf1}. The TEGR is an
alternative geometrical formulation of Einstein's general
relativity. The teleparallel geometry is determined by a set
of global orthonormal fields, or tetrad fields
$e^a\,_\mu$\cite{Weitzenbock,Einstein,Schouten}.
The action integral is determined by a particular combination
of quadratic terms in the torsion tensor. A definition for the
gravitational energy has been  established in the Hamiltonian
formulation of the TEGR\cite{Maluf2,Maluf3}.

We will evaluate the gravitational energy of the space-time
defects investigated by Puntigam and Soleng\cite{Puntigam},
and arrive at an interesting result: the gravitational energy of
space-time dislocations vanishes, whereas the gravitational
energy per unit length for space-time disclinations is finite,
and therefore the total energy is infinit. The close analogy
with solid continua is clear, since in crystals and metals
dislocations and disclinations require a small and large amount
of energy, respectively, to be created. The
interpretation of the space-time as a continuum with
microstructure has interesting consequences.

The determination of the energy-momentum tensors that generate
these defects is an important issue that can be easily
investigated in the present mathematical setting, by means of
our defect model. The emergence of squares of delta
functions has been pointed out earlier\cite{Galtsov,Letelier}
in the analysis of the chiral string (a dislocation field). Our
procedure will confirm the appearance of such terms.\par

\bigskip
\noindent Notation: space-time indices $\mu, \nu, ...$ and SO(3,1)
indices $a, b, ...$ run from 0 to 3. Latin 
indices from the middle of the alphabet indicate space indices 
according to $\mu=0,i,\;\;a=(0),(i)$. The tetrad field $e^a\,_\mu$ 
yields the definition of the torsion tensor:  
$T^a\,_{\mu \nu}=\partial_\mu e^a\,_\nu-\partial_\nu e^a\,_\mu$.
The flat space-time  metric is fixed by
$\eta_{ab}=e_{a\mu} e_{b\nu}g^{\mu\nu}= (-+++)$.        \\

\bigskip
\bigskip
\noindent {\bf II. The Lagrangian formulation of the TEGR}\par
\bigskip

The Lagrangian formulation of the TEGR is formulated in terms
of the tetrad field $e^a\,_\mu$, and is given by a sum of
quadratic terms in the torsion tensor $T^a\,_{\mu \nu}$, which
is related to the anti-symmetric part of Cartan's connection
$\Gamma^\lambda _{\mu \nu}= e^{a\lambda}\partial_\mu e_{a\nu}$.
The curvature tensor constructed out of the latter
vanishes identically. This connection defines a space with
teleparallelism, or absolute parallelism\cite{Schouten}.
The Lagrangian density is given by

$$L(e)\;=\;-k\,e\,\Sigma^{abc}T_{abc}\;-\;L_M\;,\eqno(1)$$

\noindent where $k={1\over {16\pi G}}$, $G$ is Newton's constant,
$e=det(e^a\,_\mu)$, $T_{abc}=e_b\,^\mu e_c\,^\nu T_{a \mu \nu}$ and

$$\Sigma^{abc}\;=\;{1\over 4}(T^{abc}+T^{bac}-T^{cab})+
{1\over 2}(\eta^{ac}T^b-\eta^{ab}T^c)\;.$$

\noindent $L_M$ is the Lagrangian density for matter fields.
Tetrads transform space-time into SO(3,1) indices
and vice-versa. The trace of the torsion tensor is
defined by $T_b=T^a\,_{ab}\;$.
The tensor $\Sigma^{abc}$ is defined such that

$$\Sigma^{abc}T_{abc}\;=\; {1\over 4} T^{abc}T_{abc} + 
{1\over 2}T^{abc}T_{bac}-T^aT_a\;.$$

The field equations obtained from (1) read

$${{\delta L}\over {\delta e^{a\mu}}}\;=\;
e_{a\lambda}e_{b\mu}\partial_\nu(e\Sigma^{b\lambda \nu})-
e\biggl(\Sigma^{b \nu}\,_aT_{b\nu \mu}-
{1\over 4}e_{a\mu}T_{bcd}\Sigma^{bcd}\biggr)
\;-\;{1\over {4k}}eT_{a\mu}\;=\;0\;,\eqno(2)$$

\noindent where $eT_{a\mu}=
{\delta L_M} /\delta e^{a\mu}$.
It can be shown by explicit calculations\cite{Maluf1} 
that for $T_{a\mu}=0$ these equations yield Einstein's 
equations,

$${{\delta L}\over {\delta e^{a\mu}}}\; \equiv \;{1\over 2}\,e\,
\biggl\{ R_{a\mu}(e)-{1\over 2}e_{a\mu}R(e)\biggr\}\;=0\;.$$

\noindent Therefore the energy-momentum tensor appearing in (2)
is strictly equivalent to the corresponding tensor on the
right hand side of Einstein's equations. Rewriting equation
(2) we have

$${1\over k}T_{a\mu}\;=\;
e_{a\mu}T_{bcd}\Sigma^{bcd}\;+\;
{4\over e}e_{a\lambda}e_{b\mu}\partial_\nu(e\Sigma^{b\lambda \nu})
\;-\;4\Sigma^{b\nu}\,_aT_{b\nu\mu}\;.\eqno(3)$$

The equivalence of the present teleparallel theory with
Einstein's general relativity must be clarified. First we note
that the action integral and field equations (2) remain well
defined even when $e^a\,_\mu$ is degenerate. Einstein's
equations in terms of tetrad fields also allow degenerate
tetrad solutions, which are assumed to induce a topology
change of the space-time\cite{Horowitz}. However,
equivalence of (2) with Einstein's equations in the standard
metrical form holds only for {\it non-degenerate} tetrad
configurations. We do not consider the possibility of
topology change of the space-time and therefore 
restrict our attention to non-degenerate tetrad configurations
only. We remark that the TEGR is necessarily defined
on parallelizable manifolds (in the sense of Ref.
\cite{Geroch1}), whereas the standard general relativity 
may be defined also on non-parallelizable manifolds.

The expression for the gravitational energy arises in the
Hamiltonian formulation of the TEGR\cite{Maluf1}. The energy\
enclosed by a volume $V$ of the three-dimensional space
is given by\cite{Maluf2}

$$E_g\;=\; {1\over {8\pi}}\int_V d^3x\,\partial_i(eT^i)
={1\over {8\pi}}\int_S dS_i(eT^i) \;.\eqno(4)$$

\noindent All field quantities in (4) are
restricted to the three-dimensional spacelike hypersurface;
$e$ is now the determinant of the  triads $e_{(k)j}$
and $T^i$ is the trace of the torsion tensor:
$T^i=g^{ik}T_k=g^{ik}e^{(l)j}T_{(l)jk}$,
$T_{(l)jk}=\partial_j e_{(l)k}-\partial_k e_{(l)j}$. 

This expression has been thoroughly examined in the
literature; it yields the ADM energy\cite{Maluf2}, and a value
strikingly close to the irreducible mass of the Kerr black
hole\cite{Maluf4}. It has also been applied to the analysis of
the gravitational energy of simple space-time
defects\cite{Maluf5}. However, in order to deal with metric
tensors of more intricate space-time defects we need a
suitable mathematical description of the defect.\par

\bigskip
\bigskip
\noindent {\bf III. The defect model}\par
\bigskip
Idealized space-time defects are represented by constant
parameters in the metric tensor. For instance, a space-time
dislocation is represented by the metric tensor

$$ds^2\;=\;-dt^2+dr^2+r^2\,d\phi^2+(dz+\gamma\,d\phi)^2\;,$$

\noindent for constant $\gamma$. One can safely take derivatives
of the metric components for $r\ne0$. However, for $r=0$ one has
to be cautius, since the metric tensor as well as the
corresponding tetrads are singular at $r=0$, and the
emergence of delta functions with support at the $z$ axis is
unavoidable. One way of addressing this problem is to consider
the metric tensor in cartesian coordinates\cite{Galtsov}, 
making use of distributional identities in the $(x,y)$
plane\cite{Gelfand}.

In this paper we will investigate space-time defects in
cylindrical coordinates, by allowing $\gamma$
to be a suitable function of the distance
$r=\sqrt{x^2+y^2}$, that circumvents the above mentioned
problem of derivatives in the $z$ axis, and that yields the
desired physical features. We establish the defect model by
choosing $\gamma(r)$ to be

$$\gamma(r)= \cases{\gamma_0, & if $r>r_0$\cr
                0, & if $r\le r_0$\cr}\;.\eqno(5)$$

\noindent where $\gamma_0$ is a constant measure of the defect.
As a consequence, the radial derivative of all
space-time defects in this article will be given by Dirac's
delta function,

$${d\over {dr}}\gamma(r)=
\gamma^\prime(r)=
\gamma_0\delta(r-r_0)\;.\eqno(6)$$

In the following section we will need to evaluate the integral
of the product of $\gamma(r)$ with $\gamma^\prime(r)$
(Eqs. (22) and (29)).
It can be calculated directly,

$$\int_0^{+\infty} dr\,\gamma^\prime(r)\,\gamma(r)=
{1\over 2}\int_0^{+\infty} dr{d\over{dr}}
\lbrack \gamma^2(r)\rbrack=
{1\over 2}\lbrack \gamma^2(+\infty)-\gamma^2(0)\rbrack=
{1\over 2}\gamma_0^2\;\;,\eqno(7)$$

\noindent or via integration by parts,

$$\int_0^{+\infty} dr \,\gamma^\prime (r)\,\gamma(r)=
\int_0^{+\infty} dr {d\over{dr}}\lbrack
\gamma^2(r) \rbrack-
\int_0^{+\infty} dr\,\gamma(r)
{d\over{dr}}\lbrack \gamma(r)\rbrack \;,$$

\noindent yielding the same result. 
We note that in order to obtain equation (7) it
is not necessary to assume any specific value for
$\gamma(r_0)$. The latter quantity is model dependent,
and therefore is arbitrarily defined.

For a given arbitrary function $f(r)$ we have

$$\int_0^{+\infty} dr\, f(r)\gamma^\prime(r) \gamma(r)=
{1\over 2}\int_0^{+\infty} dr\,
f(r){d\over{dr}}\gamma^2(r)$$

$$={1\over 2}\int_0^{+\infty} dr\,
{d\over{dr}}\biggl(f(r) \gamma^2(r)\biggr)-
{1\over 2} \int_0^{+\infty} dr\, {{df(r)} \over{dr}}
\gamma^2(r)\;,\eqno(8) $$

\noindent from what follows

$$\int_0^{+\infty} dr\, f(r)\gamma^\prime(r) \gamma(r)=
{1\over 2}\gamma_0^2f(r_0)\;. \eqno(9)$$

\noindent Again, we note that in order to obtain equation (9)
it it not necessary to assume any particular value for
$\gamma(r_0)$.

As a final step, we take the limit $r_0 \rightarrow 0$.
This procedure 
allows us to evaluate in a straightforward way
several expressions of energy-momentum tensors that have been
obtained in the literature by means of alternative methods.

\bigskip
\bigskip
\noindent {\bf IV. The total gravitational energy of dislocations
and disclinations}\par
\bigskip
In real crystals disclinations are highly energetic defects
(see, for instance, section  9.2 of Ref. \cite{Chaikin}), whereas
dislocations require a small amount of energy to be formed. We
will conclude that there is a similar situation in the context
of space-time defects, by applying expression (4) to the
evaluation of the gravitational energy.

We will consider the distorted space-times constructed by
Puntigam and Soleng\cite{Puntigam}. Out of the ten space-times
described by these authors we will take into account only
those whose spacelike section is time independent. The six
defects described
from order 1 through 6 are obtained by means of the Volterra
process in $\bf {R}^3$, and are related to the six degrees of
freedom of the proper group of motion of the Euclidean group
$SO(3)\otimes T(3)$. As mentioned
in Ref. \cite{Puntigam}, space {\it and} time supported
distortions have no analogy in the theory of elasticity.
Therefore we will dispense with orders 8, 9 and 10 of
Ref. \cite{Puntigam} and consider the metric tensors of order
1 through 6, as displayed in Tables 1 and 2.

\begin{table}
\caption{Space-time dislocations}
\begin{center}
\begin{tabular}{|c||r||r|}\hline
 Order & metric tensor\\ \hline
1   & $ds^2=-dt^2+dx^2+dy^2+dz^2+2{\gamma \over {r^2}}dx(xdy-ydx)+   
({\gamma\over{r^2}})^2(xdy-ydx)^2$\\  \hline
2   & $ds^2=-dt^2+dx^2+dy^2+dz^2+2{\gamma \over {r^2}}dy(xdy-ydx)+   
({\gamma\over{r^2}})^2(xdy-ydx)^2$\\  \hline
3   & $ds^2=-dt^2+dr^2+r^2\,d\phi^2+(dz+\gamma d\phi)^2$\\  \hline
\end{tabular}
\end{center}
\end{table}

According to the procedure outlined in the  previous section,
the quantities $\beta$ and $\gamma$ will be assumed
{\it a priori} as functions of the radial distance $r$
according to equation (5).

The energy expression (4) has been
obtained by requiring the time
gauge condition, which implies a Hamiltonian formulation with a
unique time scale. The tetrad fields below satisfy the time
gauge condition. All energy expressions will be evaluated
first on a cylindrical surface $S$ of radius $R>r_0$ and height
$L$; then we will make $R\rightarrow \infty$ and
$L\rightarrow \infty$.
The prime indicate derivative with
respect to the radial distance $r$.\par

\begin{table}
\caption{Space-time disclinations}
\begin{center}
\begin{tabular}{|c| |r| |r|}\hline
 Order & metric tensor\\ \hline
4  & $ds^2=-dt^2+dx^2+dy^2+dz^2$\\
{} & $-2{\beta\over{r^2}}(zdy-ydz)(xdy-ydx)
+({\beta\over{r^2}})^2(y^2+z^2)(xdy-ydx)^2$\\ \hline
5  & $ds^2=-dt^2+dx^2+dy^2+dz^2$\\
{} & $-2{\beta\over{r^2}}(zdx-xdz)(xdy-ydx)
+({\beta\over{r^2}})^2(x^2+z^2)(xdy-ydx)^2$\\ \hline
6  & $ds^2=-dt^2+dr^2+(\beta\,r)^2d\phi^2+dz^2$\\   \hline
\end{tabular}
\end{center}
\end{table}

\bigskip
\noindent {\it Order 1}\par
\bigskip
The metric tensor given in Table 1 can be rewritten in
cylindrical coordinates as

$$ds^2=-dt^2+dr^2+\sigma^2d\phi^2+2\gamma\cos\phi\,dr\,d\phi
+dz^2\;,\eqno(10)$$

\noindent where $\sigma^2=r^2+\gamma^2-2\gamma\,r\,\sin\phi$.
We obtain

$$e_{a\mu}=\pmatrix{
-1&0&0&0\cr
0 & \cos\phi & -r\sin\phi + \gamma & 0\cr
0 & \sin\phi & r\cos\phi & 0\cr
0&0&0&1\cr}\;,\eqno(11)$$

\noindent from what follows the only non-vanishing torsion
component $T_{(1)12}=\gamma^\prime\;$.
Out of the latter quantity we calculate
$T^1=-{1\over e^2}\gamma^\prime( \gamma-r\sin\phi)$,
$T^3=0$, $e=\det(e^a\,_\mu)=\det(e_{(i)j})=
(r-\gamma\,\sin\phi)$, and therefore

$$E_g={1\over {8\pi}}\int_S dS \biggl[
 -{1\over e} \gamma^\prime( \gamma - r\sin\phi)
 \biggr]\;,\eqno(12)$$

\noindent where $dS=d\phi\, dz$.
The integration is only on the cylindrical surface
determined by $r=R$ because $T^3=0$. We will analyze separately
the two terms above.

We consider first $-{1\over e}\gamma^\prime \gamma$. In view
of Eqs. (5) and (6) it can be written as

$$-{1\over e}\gamma^\prime \gamma
=-{1\over 2e}{d\over {dr}}(\gamma^2)
=-{1\over {2e}}\gamma_0^2\delta(r-r_0)\;.\eqno(13)$$

\noindent The integral of equation (15) on a cylindrical surface
$S$ of radius $R>r_0$ yields

$$-{1\over{8\pi}}\int_S d\phi\,dz\,
{1\over {2e}}\gamma_0^2\delta(r-r_0)=
-{L\over {16\pi}} \gamma_0^2 \int_{r=R}
d\phi\, {1\over e}\delta(r-r_0)=0\;.$$

Next we consider the term ${1\over e}\gamma^\prime\,r\,\sin\phi$.
We have

$${1\over e}\gamma^\prime\,r\, \sin\phi=
{1\over {\lbrack r-\gamma(r) \sin\phi\rbrack}}
\gamma_0 \delta(r-r_0)\,r\,\sin\phi=
{1\over r_0}\gamma_0\delta(r-r_0)\,r_0\,\sin\phi\;.\eqno(14)$$

\noindent The integral on $S$ of equation (14)
vanishes, what implies the vanishing of equation (12).

\bigskip
\noindent {\it Order 2}\par
\bigskip In cylindrical coordinates the metric tensor for
order 2 reads as

$$ds^2=-dt^2+dr^2+
\sigma^2d\phi^2+2\gamma\,\sin\phi\,dr\,d\phi
+dz^2\;,\eqno(15)$$

\noindent where $\sigma^2=r^2 +\gamma^2+2\gamma\,r\,\cos\phi$.
Out of the metric tensor above we can construct the set of
tetrads

$$e_{a\mu}=\pmatrix{
-1&0&0&0\cr
0 & \cos\phi & -r\sin\phi & 0\cr
0 & \sin\phi & r\cos\phi+\gamma & 0\cr
0&0&0&1\cr}\;.\eqno(16)$$

\noindent The only non-vanishing tensor component is given by
$T_{(2)12}=\gamma^\prime\;$. It is not difficult to obtain
$T^1=-{1\over e^2}\gamma^\prime( \gamma+r\cos\phi)$,
$T^3=0$ and $e=(r+\gamma\,cos\phi)$. The energy contained within
a cylindrical surface $S$ is given by

$$E_g={1\over {8\pi}}\int_S dS\biggl[
-{1\over e}\gamma^\prime (\gamma + r\cos\phi)\biggr]
\;,\eqno(17)$$

\noindent Folowing the same reasoning presented above for order
1 we easily conclude that equation (17) vanishes. We 
remark that had we assumed from the outset that
$\gamma=\gamma_0$ is a constant for any $r$, both in
orders 1 and 2, we would also arrive at a vanishing value
for the gravitational energy. The assumption
that $\gamma$ is {\it a priori} given by equation (5)  is
important in the analysis of orders 4 and 5 and
for obtaining the energy-momentum tensors.\par

\bigskip
\noindent {\it Order 3}\par
\bigskip

The simplest set of tetrad fields that yields the metric tensor
of order 3 is given by

$$e_{a\mu}=\pmatrix{
-1&0&0&0\cr
0&\cos\phi& -r\sin\phi& 0\cr
0&\sin\phi& r\cos\phi&0\cr
0&0&\gamma(r)&1\cr}\;.\eqno(18)$$

\noindent Again there is only one non-vanishing torsion tensor
component, $T_{(3)12}=\gamma^\prime\;$, that yield
vanishing values for all torsion traces
$T^i$. Consequently we arrive at $E_g=0$.\par

\bigskip
\noindent {\it Order 4}\par
\bigskip

In cylindrical coordinates the metric tensor in Table 2 is
rewritten as

$$ds^2=-dt^2+dr^2+\sigma^2d\phi^2+
dz^2-2\beta z\,\sin\phi\,dr\,d\phi+
2\beta r\,\sin\phi\,d\phi\,dz\;,\eqno(19)$$

\noindent where $\sigma^2=e^2+(\beta r\,\sin\phi)^2+
(\beta z\,\sin\phi)^2$, and $e=(r-\beta z\,\cos\phi)$. The tetrad
representation of equation (19) reads

$$e_{a\mu}=\pmatrix{
-1&0&0&0\cr
0&\cos\phi&-r\sin\phi&0\cr
0&\sin\phi&r\cos\phi-\beta z&0\cr
0&0&\beta r\,\sin\phi&1\cr} \;.\eqno(20)$$

\noindent The non-vanishing torsion tensor components are
$T_{(2)12}=-z\beta^\prime ,\;\;
T_{(2)23}=\beta ,\;\;
T_{(3)12}=(\beta+r\beta^\prime)\sin\phi\;$.
We obtain $eT^1=-{1\over e}\beta^\prime z(\beta z-
r\cos\phi)$ and $eT^3=-\beta \cos\phi+
{1\over e}\beta^\prime \beta z r\,\sin^2\phi$.

Substituting these quantities in expression (4) we find
$E_g=E_{g1}+2E_{g2}$, where

$$E_{g1}={1\over {8\pi}}\int_{-{L\over 2}}^{L\over 2}dz
\int_0^{2\pi}d\phi\biggl[ -{1\over e}\beta^\prime z(\beta z-
r\cos\phi)\biggr]\;,\eqno(21)$$

$$E_{g2}={1\over {8\pi}}\int_0^R dr
\int_0^{2\pi}d\phi \biggl[ -\beta \cos\phi+
{1\over e}\beta^\prime \beta z r\,\sin^2\phi \biggr]
\;.\eqno(22)$$

\noindent $E_{g1}$ is evaluated at the surface of constant
radius $R>r_0$ and height $L$,
and $2E_{g2}$ corresponds to the integrations
both at $z={L\over 2}$ and at $-{L\over 2}$. Repeating the
arguments that led to equations (13) and (14) we conclude that
the first and second terms in $E_{g1}$ vanish under
integration.

The first term in $E_{g2}$ also vanishes under integration
with respect to $\phi$. The second term yields a non-vanishing
result. By making $z={L\over 2}$ and considering

$$\beta^\prime(r)\beta(r)=
{1\over 2}{d\over {dr}}\lbrack \beta(r)\rbrack^2=
{1\over 2}\beta_0^2\delta(r-r_0)\;,\eqno(23)$$

\noindent we find

$$E_{g2}={1\over {8\pi}}{L\over 2}\int_0^{2\pi}d\phi \sin^2\phi
\int_0^R dr {1\over 2}(\beta_0)^2\delta(r-r_0)
\biggl[ {r\over{r-\beta(r) {L\over 2}\cos\phi}}
\biggr]$$

$$={L\over {32\pi}}(\beta_0)^2 \int_0^{2\pi}
d\phi \sin^2\phi={L\over{32}}(\beta_0)^2 \;.\eqno(24)$$

\noindent Note that the use of equation (23) in this calculation
is essentially equivalent to  using equation (9).
Thus the total energy contained within a cylinder of
height $L$ is given by

$$E_g = L\biggl({\beta_0\over 4}\biggr)^2 \;.\eqno(25)$$

\noindent Therefore $E_g$ diverges in the limit
$L\rightarrow \infty$.

\bigskip
\noindent {\it Order 5}\par
\bigskip

The calculations here are similar to those of order 4.
The metric tensor of order 5 in cylindrical coordinates is
written as

$$ds^2=-dt^2+dr^2+\sigma^2d\phi^2+dz^2-
2\beta z\,\cos\phi\, dr\,d\phi+
2\beta r\,\cos\phi\,d\phi dz\;,\eqno(26)$$

\noindent where $\sigma^2= e^2+(\beta r\,\sin\phi)^2+
(\beta z\,\sin\phi)^2$, and $e=(r+\beta z\,\sin\phi)$.
This space-time disclination can be described by the set of
tetrad fields

$$e_{a\mu}=\pmatrix{
-1&0&0&0\cr
0&\cos\phi&-r\sin\phi-\beta z&0\cr
0&\sin\phi&r\cos\phi&0\cr
0&0&\beta r\cos\phi&1\cr}\;.\eqno(27)$$

\noindent Three components of the torsion tensor are
non-vanishing: 
$T_{(1)12}=-z\beta^\prime ,\;\;
T_{(1)23}=\beta ,\;\;
T_{(3)12}=(\beta+r\beta^\prime)\cos\phi\;$.
It is not difficult to obtain
$eT^1=-{1\over e}\beta^\prime z(\beta z+r\sin\phi)$ and
$eT^3=-\beta \sin\phi +
{1\over e}\beta^\prime \beta z r \cos^2\phi$. The total
gravitational energy is given by $E_g=E_{g1}+2E_{g2}$, where

$$E_{g1}={1\over {8\pi}}
\int_{-{L\over 2}}^{L\over 2}dz
\int_0^{2\pi}d\phi\biggl[
-{1\over e}\beta^\prime z(\beta z+r\sin\phi)
\biggr]\;,\eqno(28)$$

$$E_{g2}={1\over {8\pi}}\int_0^R dr
\int_0^{2\pi}d\phi \biggl[ -\beta \sin\phi +
{1\over e}\beta^\prime \beta z r \cos^2\phi
\biggr]\;.\eqno(29)$$

\noindent $E_{g1}$ is evaluated at the surface of constant 
radius $R>r_0$, and $2E_{g2}$ corresponds to the
integrations at $z=\pm {L\over 2}$. Following
precisely the same steps that led to expression (24) we find
that $E_{g1}$ and the integration of the first term of
$E_{g2}$ vanish, and

$$E_g=2E_{g2}= 2{L\over {32\pi}}(\beta_0)^2 \int_0^{2\pi}
d\phi \cos^2\phi=L \biggl( {\beta_0\over 4}\biggr)^2
\;,\eqno(30)$$

\noindent from what we conclude that the total energy $E_g$
diverges in the limit $L \rightarrow \infty$.\par

\bigskip
\noindent {\it Order 6}\par
\bigskip
The metric tensor for order 6 describes the usual cosmic string.
The simplest realization of the tetrad fields that yield the
metric tensor in Table 2 is 

$$e_{a\mu}=\pmatrix{
-1&0&0&0\cr
0&\cos\phi&-\beta\sin\phi &0\cr
0&\sin\phi&\beta\cos\phi &0\cr
0&0&0&1\cr}\;.\eqno(31)$$

\noindent The determinant is $e=\beta r$ and the two
non-vanishing torsion tensor components are
$T_{(1)12}=(1-\beta-r\beta^\prime)\sin\phi,\;\;
T_{(2)12}=-(1-\beta-r\beta^\prime)\cos\phi\;.$
We obtain $eT^1=(1- \beta-r\beta^\prime )$
and $T^3=0$. In this order the defect function is defined by

$$\beta(r)= \cases{1, & if $r \le r_0$\cr
                \beta_0, & if $r > r_0$\cr}\;,$$

\noindent from what follows $\beta^\prime(r)
=-(1-\beta_0)\delta(r-r_0)$. For $r_0 \rightarrow 0$
we obtain

$$E_g={1\over {8\pi}}\int_{-{L\over 2}}^{L\over 2}dz
\int_0^{2\pi}d\phi (1-\beta-r\beta^\prime)=
{L\over 4}(1-\beta_0)\;,\eqno(32)$$

\noindent which is the well-known expression for the energy
of the cosmic string. Again the energy diverges
in the limit $L\rightarrow \infty$.
Therefore for the three disclinations we obtain a finite
value for the gravitational energy per unit
length of the defect.\par

\bigskip
\bigskip
\noindent {\bf V. Energy-momentum tensors}\par
\bigskip
An important issue concerning the physical viability of the
space-time defects considered here is the determination of the
energy-momentum tensors that generate the defects.
We antecipated in section I that in previous investigations of
some of these defects (orders 3 and 6)
there appeared linear {\it and} quadratic terms
in the delta function. In this section we will present the
energy-momentum tensors corresponding to orders 1 through 6,
by  means of expression (3), and will also arrive at
expressions containing squares of the delta function.
We remark that it is possible to demonstrate by explict
calculations that expression (3) is symmetric, i.e., 
$T_{\mu\nu}=e_a\,^\mu T_{a\mu}=T_{\nu\mu}$.

The calculations are quite lengthy, but otherwise
straightforward. We need to evaluate all components
$T_{\lambda\mu\nu}=e^a\,_\lambda T_{a\mu\nu}$ and of the
tensor $\Sigma^{\mu\nu\lambda}$ defined in section II. All
components of $T_{a\mu\nu}$ were evaluated in the previous
section. In what follows we will just present the final
expressions of the non-vanishing components of 
$T_{\mu\nu}$ exactly as they arise from the calculations
(except for orders 4 and 5; see below). The expressions
will be given in terms of the
functions $\beta(r)$, $\gamma(r)$ and their
derivatives. Therefore it will be understood that 
$\beta^\prime(r)=\beta_0\delta(r-r_0)$ (for order 6,
$\beta^\prime(r)=-(1-\beta_0)\delta(r-r_0)$) and
$\gamma^\prime(r)=\gamma_0\delta(r-r_0)$, for arbitrarily
small $r_0$.\par

\bigskip
\noindent {\it Order 1}\par

$${{T_{00}}\over k}=-{{T_{33}}\over k}=
-{2\over e}(\gamma-r\sin\phi) \partial_r\biggl[
{\gamma^\prime \over e}\biggr]
- {2\over e^2}  (\gamma^\prime)^2+
{2\over e}  \gamma \gamma^\prime \cos^2\phi \;.$$

\bigskip
\noindent {\it Order 2}\par

$${{T_{00}}\over k}=-{{T_{33}}\over k}=
-{2\over e} (\gamma+r\cos\phi) \partial_r\biggl[
{\gamma^\prime \over e}\biggr]
-{2 \over e^2}(\gamma^\prime)^2 -
{2\over e} \gamma \gamma^\prime \sin^2\phi \;.$$

\bigskip
\noindent {\it Order 3}\par

$${{T_{00}}\over k}=-{{T_{11}}\over k}=
-{{(\gamma^\prime)^2}\over{2e^2}}\;,\;\;\;\;\;
{{T_{22}}\over k}=
-2\gamma\,e\,\partial_r\biggl[
{\gamma^\prime\over e}\biggr]-{{(\gamma^\prime)^2}\over 2}+
\gamma^2\biggl[ {{3(\gamma^\prime)^2}\over {2e^2}}\biggr]\;,$$

$${{T_{23}}\over k}=
-e\,\partial_r\biggl[
{\gamma^\prime\over e}\biggr]+
\gamma\biggl[ {{3(\gamma^\prime)^2}\over {2e^2}}\biggr]\;,
\;\;\;\;\;{{T_{33}}\over k}=
{{3(\gamma^\prime)^2}\over {2e^2}}
\;.$$

\bigskip
\noindent {\it Order 4}\par

$${{T_{00}}\over k}=
-{2\over e} z(\beta z-r\,\cos\phi)\,
\partial_r \biggl[{\beta^\prime\over e}\biggr]
+{2\over e^3}\beta \beta^\prime(z^2+r^2)\,\sin^2\phi
\;,$$

$${{T_{11}}\over k}=-{2\over e^2}\beta \beta^\prime r\,
\sin^2\phi\;,$$

$${{T_{12}}\over k}={1\over e} \beta^2 zr^2\sin^3\phi
\partial_r\biggl[{\beta^\prime \over e}\biggr]+
{1\over e}\beta \beta^\prime r\,\sin\phi\cos\phi
\biggl( 1+{r^2\over e^2}\biggr)\;,$$

$${{T_{13}}\over k}={1\over e}\beta zr\,\sin^2\phi\,
\partial_r\biggl[{\beta^\prime \over e}\biggr]+
{1\over e^2}\beta^\prime (3e\,\cos\phi-\beta z)
-{2\over e^2}\beta \beta^\prime z\,\cos^2\phi
-{1\over e^3}\beta \beta^\prime zr\,\sin^2\phi
\;,$$

$${{T_{22}}\over k}=-2\beta r^3\sin^2\phi\,
\partial_r\biggl[ {\beta^\prime \over e}\biggr]
-{6\over e^2}\beta\beta^\prime r^3\sin^2\phi
\;,$$

$${{T_{23}}\over k}={1\over e}r\,\sin\phi(\beta^2z^2-r^2)
\partial_r \biggl[{\beta^\prime\over e}\biggr]
-{2\over e} \beta \beta^\prime z \sin\phi\cos\phi
+{3\over e^2}\beta^\prime \sin\phi(\beta zr\,\cos\phi-
r^2),\;$$

$${{T_{33}}\over k}={2\over e}z(\beta z-r\cos\phi)
\partial_r \biggl[{\beta^\prime\over e}\biggr]
-{2\over e^3}\beta \beta^\prime z^2\sin^2\phi
\;.$$

\bigskip
\noindent {\it Order 5}\par

$${{T_{00}}\over k}=-{2\over e}
z(\beta z+ r\,\sin\phi)\,
\partial_r \biggl[{\beta^\prime\over e}\biggr]
+{2\over e^3}\beta \beta^\prime(z^2+r^2)\cos^2\phi
\;,$$

$${{T_{11}}\over k}=-{2\over e^2}\beta \beta^\prime r\,
\cos^2\phi\;,$$

$${{T_{12}}\over k}={1\over e}\beta^2 z r^2\cos^3\phi\,
\partial_r\biggl[ {\beta^\prime \over e}\biggr]-
{1\over e}\beta \beta^\prime r\,\sin\phi\cos\phi
\biggl(1+{r^2\over e^2}\biggr)\;,$$

$${{T_{13}}\over k}={1\over e}\beta zr\,\cos^2\phi\,
\partial_r\biggl[{\beta^\prime \over e}\biggr]-
{1\over e^2}\beta^\prime  (3e\,\sin\phi+\beta z)
+{2\over e^2}\beta \beta^\prime z\,\sin^2\phi
-{1\over e^3}\beta \beta^\prime zr\,\cos^2\phi
\;,$$

$${{T_{22}}\over k}=-2\beta r^3\cos^2\phi\,
\partial_r\biggl[ {\beta^\prime \over e}\biggr]
-{6\over e^2}\beta\beta^\prime r^3\cos^2\phi
\;,$$

$${{T_{23}}\over k}={1\over e}r\,\cos\phi(\beta^2z^2-r^2)
\partial_r \biggl[{\beta^\prime\over e}\biggr]
+{2\over e} \beta \beta^\prime z \sin\phi\cos\phi
-{3\over e^2}\beta^\prime \cos\phi(\beta zr\,\sin\phi-
r^2)\;,$$

$${{T_{33}}\over k}={2\over e}z(\beta z+r\sin\phi)
\partial_r \biggl[{\beta^\prime\over e}\biggr]
-{2\over e^3}\beta \beta^\prime z^2\cos^2\phi
\;,$$

\bigskip
\noindent {\it Order 6}\par

$${{T_{00}}\over k}=-{{T_{33}}\over k}=
{2\over e}\partial_r \biggl[1-\beta-r\beta^\prime\biggr]
\;.$$

\bigskip
\noindent For each order $e$ is the determinant obtained in
the previous section.

The full expressions for the $T_{\mu\nu}$ components of orders
4 and 5 are very intricate. We have just presented the terms
that are linear in $\beta^\prime$. These are
the relevant terms to be considered, according to the suggestion
pointed out in Refs. \cite {Galtsov,Letelier}. In these latter
references it is argued that the terms quadratic in
the delta function are devoid of physical meaning, and that
they should be ignored.

The terms {\it linear} in the defects 
in the expressions above for orders 3 and
6 are in total agreement with the corresponding expressions
obtained in Refs. \cite{Galtsov,Letelier}, except that in these
references the calculations were carried out in cartesian
coordinates (the quadratic terms were not presented in
\cite{Galtsov,Letelier}).\par

\bigskip
\bigskip
\noindent {\bf VI. Discussion}\par
\bigskip

The results of section IV indicate that there is an analogy
between space-time defects and defects in continuum material
systems. In the latter, dislocations and disclinations are low
and high energy defects, respectively. For the idealized
defects considered in this work we obtained an analogous result.

Deformations of metals are explained entirely in terms of
dislocations. It the analysis of the plastic properties of
metals it is made no reference to disclinations\cite{Smallman}.
Deformation of such materials is due to the mobility
and multiplication of dislocations. However, deformations of the
space-time that would give rise to the formation of matter
structure in the universe are presently described by
disclinations similar to order 6.
The value of the energy per unit length (32)
for the usual cosmic string is well-known from previous
studies\cite{Vilenkin}. To our knowledge, energies 
(25) and (30) have not been calculated so far. As for
the dislocations, all have vanishing total energy.

The teleparellel geometry is determined by the choice of a
global orthonormal set of tetrad fields. It is established by
just declaring a particular orthonormal frame to be the set of
tetrad fields for the space-time\cite{Nester}. All tetrad fields
considered here yield vanishing torsion tensor by requiring
the physical parameters $\beta_0$ and $\gamma_0$ to
vanish. Alternatively, if we require the latter parameters to
vanish, then in cartesian coordinates all tetrad fields reduce
to

$$e^a\,_\mu(t,x,y,z)=\delta^a_\mu\;.\eqno(33)$$

\noindent It is legitimate
to ask whether different choices for $e_{a\mu}$, in the six
cases above, would render different results for the energy.
The answer is that the {\it total} energy does not depend on
the particular choice of an orthonormal frame, provided the
tetrad fields are related to each other via a local SO(3,1)
transformation such that the transformation matrices satisfy
appropriate boundary conditions (section VI of Ref.
\cite{Maluf6}). The transformed tetrad fields must also reduce
to the form given by equation (33) for vanishing physical
parameters. In order to establish the
above mentioned analogy it suffices to evaluate the total
gravitational energy of the space-time.

Expression (4) for the gravitational energy was obtained from
the Hamiltonian formulation where the time gauge condition for
the tetrad field ($e_{(i)}\,^0=0$)
was imposed from the outset. If we consider
the Hamiltonian formulation of the TEGR without fixing the time
gauge condition, there arises a total divergence
in the Hamiltonian constraint that plays a role similar to
expression (4), and that is identified with the energy-momentum
of the gravitational field\cite{Maluf7}. However, it is possible
to prove\cite{prep} that by requiring $e_{a\mu}$ to satisfy
({\it a posteriori}) the time gauge condition the latter
expression for the gravitational energy
exacty coincides with expression (4).

\bigskip
\noindent {\bf Acknowledgements}\par
\noindent A. G. is supported by CAPES, Brazil.\par

\end{document}